\documentclass[pra,twocolumn,showpacs,preprintnumbers,amsmath,amssymb]{revtex4}

\usepackage{bm}
\usepackage{bbm}
\usepackage{amssymb}
\usepackage{amsfonts}

\usepackage{epsfig,graphicx}
\usepackage{amstext}
\usepackage{amsmath}
\usepackage{graphicx}
\usepackage{times}
\usepackage{txfonts}
\usepackage{dcolumn}
\usepackage{color}

\begin{document}

\title{Protecting Quantum Fisher Information in Correlated Quantum Channels}

\author{Ming-Liang Hu}
\email{mingliang0301@163.com}
\affiliation{School of Science, Xi'an University of Posts and Telecommunications, Xi'an 710121, China}

\author{Hui-Fang Wang}
\affiliation{School of Science, Xi'an University of Posts and Telecommunications, Xi'an 710121, China}

\begin{abstract}
Quantum Fisher information (QFI) has potential applications in
quantum metrology tasks. We investigate QFI when the consecutive
actions of a quantum channel on the sequence of qubits have partial
classical correlations. Our results showed that while the
decoherence effect is detrimental to QFI, effects of such classical
correlations on QFI are channel-dependent. For the Bell-type probe
states, the classical correlations on consecutive actions of the
depolarizing and phase flip channels can be harnessed to improve
QFI, while the classical correlations in the bit flip and bit-phase
flip channels induce a slight decrease of QFI. For a more
general parameterization form of the probe states, the advantage of
using initial correlated system on improving QFI can also be
remained in a wide regime of the correlated quantum channels.
\end{abstract}


\maketitle

\section{Introduction} \label{sec:1}
In quantum estimation theory, one of the fundamental tasks is to
estimate the values of the unobservable parameters in the labeled
quantum system based on the set of measurement data \cite{metro1,
metro2,metro3}. To improve the precision of such an estimation and
to approach asymptotically the quantum limit of the estimation
accuracy remain the main pursuits of this task. In this context,
many relevant works have been done in the past years, among which
the quantum Fisher information (QFI) was shown to be an essential
quantity as the Cram\'{e}r-Rao theorem shows that its inverse places
a fundamental limit on the variance of the estimator
\cite{Rao1,Rao2}. Due to this practical application, QFI has been
studied from different aspects, with much notable progresses being
achieved. For example, it has been adopted to derive a statistical
generalization of the Heisenberg uncertainty relation \cite{Luosl}.
Its measurement with finite  precision was demonstrated in
\cite{deqfi}, while its interconnections with quantum correlations
\cite{qfi-qc1,qfi-qc2,qfi-qc3,qfi-qc4} and quantum coherence
\cite{qfi-co1,qfi-co2, qfi-co3} have also been identified. Moreover,
there are several other works focused on studying properties of QFI
in certain explicit physical systems \cite{spin,Dicke,atom} and the
noninertial frames \cite{reff1,reff2}.

When implementing the parameter estimation tasks in laboratory, the
quantum system for which the probe state is encoded in will
unavoidably interacts with its surrounding environments. This may
induce rapid decay of the quantum coherence of the state \cite{deco}
as well as its quantum correlations if the state is of bipartite or
multipartite \cite{QE,QD,robu,Hu,Hulian}. Of course, such an
unwanted interaction is also detrimental to QFI in most cases
\cite{deco1,deco2,deco3,deco4,deco5,prot1,prot2}, and remains a
bottleneck restricting our ability to approach the quantum limit of
the estimation accuracy. So it is of practical significance to seek
efficient ways for protecting QFI of a system in noisy environments
\cite{PhysRep}.

Theoretically, one can regard all the unwanted effect caused by the
inevitable interaction between the principle system and its
surrounding environments as the sources of noises. In most cases,
such a noisy effect can be characterized by a quantum channel
transforming the input state into the output one. During this
transformation process, the channel may retain partial memory about
the history of its action on the sequence of qubits passing through
it \cite{Nielsen}. In general, there are two origins of the memory
effects, i.e., the one occurs during the time evolution of each
qubit due to the temporal correlations, and the one occurs between
consecutive actions of the channel \cite{cc0}. The former memory
effect may induce damped oscillations of quantum correlations such
as entanglement and quantum discord \cite{QE,QD,robu,Hu,Hulian},
while positive role of the latter memory effect on enhancing quantum
coherence of two-qubit states \cite{Zhouw} and reducing the entropic
uncertainty of two incompatible observables \cite{eur} have also
been observed.

In this paper, we examine how the memory effect caused by the
classical correlations between the consecutive actions of a channel
affects QFI. We will show that while the decoherence effect
induces rapid decay of the QFI, such a memory effect can be
harnessed to delay evidently this decay. This result reveals
distinct effects of the correlated and independent actions of a
channel on the sequence of qubits on one hand, and on the other
hand, is useful for improving the precision of parameter estimation
of an open system. The structure of this paper is as follows. In
Sec. \ref{sec:2} we recall briefly some preliminaries related to QFI
and correlated quantum channels. Then in Sec. \ref{sec:3}, we
investigate in detail influence of the correlated depolarizing, bit
flip, bit-phase flip, and phase flip channels on QFI. Finally, Sec.
\ref{sec:4} is a conclusion remark of this work.

\section{Preliminaries} \label{sec:2}
To begin with, we first recall briefly the notion of QFI and its
calculation based on the spectral decomposition of the density
operator. To define the QFI, we consider a probe state described by
$\rho_\theta$ for which $\theta$ is an unobservable parameter, then
the QFI with respect to $\theta$ can be written as \cite{metro2}
\begin{equation}\label{eq2a-1}
 F_\theta= \mathrm{Tr}(L_\theta^2\rho_\theta),
\end{equation}
where $L_\theta$ is the symmetric logarithmic derivative determined
by $\partial_\theta\rho_\theta= (L_\theta \rho_\theta +\rho_\theta
L_\theta)/2$.

By decomposing the density operator as $\rho_{\theta}=\sum_i
\lambda_i |\psi_i\rangle\langle\psi_i|$, with $\lambda_i$ being the
eigenvalues of $\rho_{\theta}$ and $|\psi_i\rangle$ the
corresponding eigenstates, an analytical solution of QFI was
derived as \cite{QFI,Liuj}
\begin{equation}\label{eq2a-2}
 F_\theta= \sum_i \frac{(\lambda'_{i})}{\lambda_{i}}^{2}
           +\sum_i {\lambda_{i}}{F_{\theta,i}}
           -\sum_{i\neq j} \frac{8 \lambda_i \lambda_j}{\lambda_i+\lambda_j}
           |\langle \psi'_i | \psi_j \rangle|^2,
\end{equation}
where we have denoted by
\begin{equation}\label{eq2a-3}
 \lambda'_i=\partial_\theta\lambda_i,~
 |\psi'_i\rangle = \partial_\theta| \psi_i\rangle, ~
 F_{\theta,i}= 4(\langle\psi'_i|\psi'_i\rangle-|\langle\psi'_i|\psi_i\rangle|^2).
\end{equation}
Obviously, $F_\theta$ is only determined by the support set of
$\rho_\theta$ and is not affected by the eigenstates outside this
support set. The first term in Eq. \eqref{eq2a-2} is determined only
by the eigenvalues of $\rho_\theta$ and is the classical
contribution, while the second and third terms can be regarded as
the quantum contribution.

An application of QFI is to describe the precision of parameter
estimation \cite{metro1,metro2,metro3}. For instance, the quantum
Cram\'{e}r-Rao theorem \cite{cramer-rao} shows that the minimum
allowed variance of the unbiased estimator $\tilde{\theta}$ for
$\theta$ is bounded by
\begin{equation}\label{c_l1}
  \mathrm{Var}(\tilde{\theta}) \geqslant \frac{1}{MF_\theta},
\end{equation}
where $\mathrm{Var}(\tilde{\theta})= \langle\tilde{\theta}^2
\rangle_{\rho_\theta}- \langle \tilde{\theta}\rangle^2_{\rho_\theta}$ ,
and $M\gg 1$ represents the times of repeated measurements.

Next, we introduce the correlated channel model. By taking $\rho_0$
as the input of the channel $\mathcal {E}$, then the output state
after the system traversing the channel can be written as $\rho=
\mathcal{E}(\rho_0)$. In the operator-sum representation, we have
$\mathcal{E}(\rho_0)= \sum_n K_n \rho_0 K_n^\dagger$, with $\{K_n\}$
being the Kraus operators describing actions of $\mathcal {E}$ on
$\rho_0$ and they satisfy the completely positive and trace
preserving relation $\sum_n K_n^\dagger K_n= \openone$
\cite{Nielsen}. We focus on the Pauli channel and the $N$-qubit
$\rho_0$, then the output state simplifies to
\begin{equation}\label{eq2b-1}
 \rho= \sum_{i_1 \cdots i_N} p_{i_1 \cdots i_N}
       (\sigma_{i_1}\otimes \cdots \otimes\sigma_{i_N})\rho_0
       (\sigma_{i_1} \otimes\cdots \otimes\sigma_{i_N}),,
\end{equation}
where $\sigma_0= \openone_2$ is the identity operator and
$\sigma_{1,2,3}$ are the Pauli operators. If $\mathcal {E}$ acts
independently and identically on each of the qubits traversing it,
the joint probability will be given by $p_{i_1 \cdots i_N}=p_{i_1}
\cdots p_{i_N}$, with $p_i\geqslant 0$ and $\sum_i p_i=1$. Such a
model describes the situation for which $\mathcal {E}$ has no memory
on the history of its actions on the sequence of qubits.

In practice, the independent actions of the channel on those qubits
traversing it is only a limiting case. When two or more qubits
traverse the channel subsequently with short time interval, the
channel may retain partial memory about its action on these qubits
\cite{cc1,cc2,cc3}. Macchiavello and Palma \cite{cc1} proposed to
describe such kind of memory effect by the classical correlations
between consecutive uses of the channel, with the joint probability
distribution function being given by
\begin{equation}\label{eq2b-2}
 p_{i_1 \cdots i_N}= p_{i_1}p_{i_2|i_1} \cdots p_{i_N|i_{N-1}},
\end{equation}
where $p_{i_n|i_{n-1}}= (1-\mu)p_{i_n}+\mu\delta_{i_ni_{n-1}}$, with
$\delta_{ij} =1$ for $i=j$ and $\delta_{ij} =0$ for $i\neq j$.
Moreover, the parameter $\mu$ characterizes strength of the
classical correlations. $\mu=0$ and 1 correspond respectively to the
uncorrelated and fully correlated channels, while the intermediate
value $0<\mu<1$ corresponds to a general classically correlated
channel.

The completely positivity and trace preserving of the channel
requires $\sum_i p_i=1$. There are many quantum channels satisfying
this requirement \cite{Nielsen}. We will focus on some paradigmatic
instances of them. They are the depolarizing, bit flip, bit-phase flip,
and phase flip channels which can be described by a parameter
$p\in[0,1]$. For the depolarizing channel, the probability
distribution function is given by
\begin{equation}\label{eq2b-3}
 p_0^{\alpha_0}=1-p, ~
 p_{1,2,3}^{\alpha_0}=\frac{p}{3},
\end{equation}
and for the bit flip ($\alpha_1$), bit-phase flip ($\alpha_2$), and
phase flip ($\alpha_3$) channels, they are given by
\begin{equation}\label{eq2b-4}
 p_0^{\alpha_k}=1-p, ~
 p_k^{\alpha_k}=p, ~
 p_{i,j\neq k}^{\alpha_k}=0.
\end{equation}

By combining these expressions with Eqs. \eqref{eq2b-1} and
\eqref{eq2b-2}, one can obtain the output state for any input state.

\section{QFI in correlated quantum channels} \label{sec:3}
We now begin our discussion about how the classical correlations
between consecutive uses of the channel affecting QFI. We take the
following two-qubit state
\begin{equation}\label{eq3-1}
 |\Phi^+\rangle= \cos\theta|00\rangle+ e^{i\varphi}\sin\theta |11\rangle,
\end{equation}
as the input of the correlated quantum channels, where $\theta$ and
$\varphi$ are unobservable parameters to be estimated. We will show
explicitly that the QFI depends strongly on the type of quantum
channels as well as strengths of the decoherence parameter and the
classical correlations.

\subsection{The correlated depolarizing channel} \label{sec:3a}
We first consider the case of correlated depolarizing channel. For
the input state $|\Phi^+\rangle$, the nonzero elements of
the density operator for the output state can be obtained as
\begin{equation}\label{eq3a-1}
\begin{aligned}
  & \rho_{11}= A\cos^{2}{\theta}+C\sin^{2}{\theta},~
    \rho_{44}=1-\rho_{11}-2B, \\
  & \rho_{22,33}=B,~
    \rho_{14}=\rho_{41}^*=(De^{-i\varphi}+E e^{i\varphi})\sin\theta\cos\theta,
\end{aligned}
\end{equation}
where the corresponding parameters are given by
\begin{equation} \label{eq3a-2}
 \begin{aligned}
  & A=(1-\eta)(1-\eta+\eta\mu), \\
  & B=\eta(1-\eta)(1-\mu), ~
    C=\eta^2+\eta(1-\eta)\mu, \\
  & D=(1-2\eta)^2+(3-4\eta)\eta\mu,~
    E=\eta\mu,
 \end{aligned}
\end{equation}
and here we have denoted by $\eta=2p/3$.

\begin{figure}[h]
\centering
\resizebox{0.45 \textwidth}{!}{%
\includegraphics{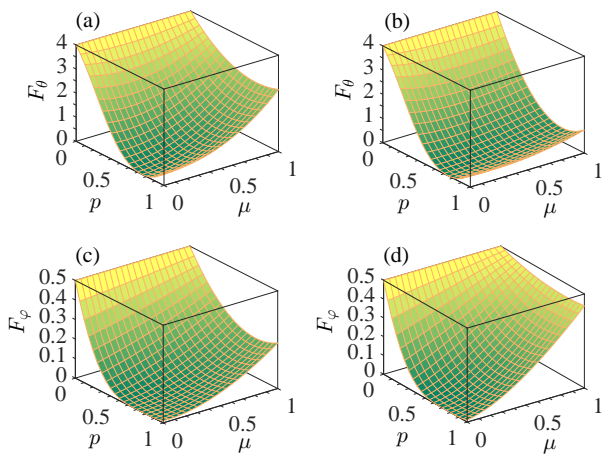}}
\caption{The $(p,\mu)$ dependence of $F_\theta$ and $F_\varphi$ for
the input state $|\Phi^+\rangle$ and the correlated depolarizing
channel. The parameters are $(\theta,\varphi)=(\pi/8,\pi/6)$ [(a) \&
(c)] and $(\pi/8,\pi/3)$ [(b) \& (d)].}\label{fig:1}
\end{figure}

For this output state $\rho$, one can derive analytically its
eigenvalues $\{\lambda_i\}$, its eigenvectors $\{|\psi_i\rangle\}$, and
their partial derivatives with respect to both $\theta$ and $\varphi$
(see Appendix \ref{sec:A}). Then one can obtain directly the QFI by
substituting these formulae into Eq. \eqref{eq2a-2}. We do not list
their expressions here due to their complexity. Instead, we performed
numerical calculation and showed in Fig. \ref{fig:1} the $(p,\mu)$
dependence of $F_\theta$ and $F_\varphi$.

It can be seen from the plots in Fig. \ref{fig:1} that with the
increasing strength of the decoherence parameter $p$, both
$F_\theta$ and $F_\varphi$ first decrease from their maxima to
certain minimum values when $p$ reaches the critical point $0.75$,
and then turn to be increased to some finite values smaller than
those at $p=0$. This shows that the decoherence effect of the
depolarizing channel is detrimental to QFI, and as a result, may
reduce its usefulness for potential quantum tasks including
parameter estimation. However, when considering the $\mu$ dependence
of the QFI, one can note from Fig. \ref{fig:1}(a) and (b) that in
the regions of relative small $p$, $F_\theta$ first decreases to
certain minimum values and then turn to be increased with the
increasing strength of $\mu$ and reaches certain maximum values
larger than those at $\mu=0$, while in the regions of large $p$, it
increases monotonically in the whole region of $\mu$. Different from
$F_\theta$, one can see from Fig. \ref{fig:1}(c) and (d) that
$F_\varphi$ is always a monotonic increasing function of $\mu$ for
any $p>0$. Hence, effects of the classical correlation between
consecutive actions of the channel on QFI is determined by both the
strength of the decoherence parameter and the parameter to be
estimated.

The above results show that the QFI can be noticeably enhanced in a
wide region of the decoherence parameter due to the correlated
actions of the depolarizing channel. This will be useful for its
practical applications including parameter estimation in noisy
environments. This is because the large value of QFI corresponds to
a small value of the variance of the estimator, hence gives an
improved estimation precision of the unobservable parameters
\cite{cramer-rao}.

\subsection{The correlated bit flip and bit-phase flip channels} \label{sec:3b}
Next, we consider the case of two qubits traversing the classically
correlated bit flip channel. The nonzero elements of the density
operator for the output state can be obtained as
\begin{equation}\label{eq3b-1}
 \begin{aligned}
  & \rho_{11}= x\cos^{2}{\theta}+z\sin^{2}{\theta},~
    \rho_{22,33}=y, \\
  & \rho_{44}=1-\rho_{11}-2y, ~
    \rho_{23,32} =y\sin2\theta\cos\varphi, \\
  & \rho_{14}=\rho_{41}^*=(xe^{-i\varphi}+ze^{i\varphi})\sin\theta\cos\theta,
 \end{aligned}
\end{equation}
where $x$, $y$, and $z$ can be obtained respectively from $A$,
$B$, and $C$ in Eq. \eqref{eq3a-2} by substituting the parameter
$\eta=2p/3$ with $\eta=p$. For this output state, though its
explicit form is a little more complicated than that of Eq.
\eqref{eq3a-1}, one can still derive analytical solutions of its
eigenvalues $\{\lambda_i\}$, its eigenvectors $\{|\psi_i\rangle\}$,
and the corresponding partial derivatives of them with respect to
both $\theta$ and $\varphi$ (see Appendix \ref{sec:B}). Hence one
can also derive straightforwardly analytical expressions of
$F_\theta$ and $F_{\varphi}$.

\begin{figure}[h]
\centering
\resizebox{0.45 \textwidth}{!}{%
\includegraphics{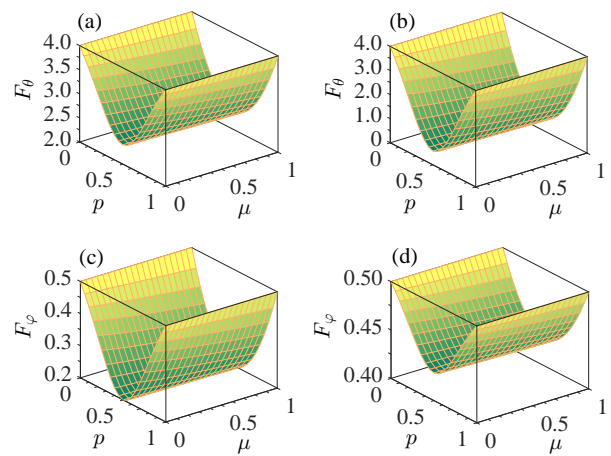}}
\caption{The $(p,\mu)$ dependence of $F_\theta$ and $F_\varphi$ for
the input state $|\Phi^+\rangle$ and the correlated bit flip
channel. The parameters are $(\theta,\varphi)=(\pi/8,\pi/6)$ [(a) \&
(c)] and $(\pi/8,\pi/3)$ [(b) \& (d)].}\label{fig:2}
\end{figure}

By choosing the same parameters as those in Fig. \ref{fig:1}, we
performed numerical calculation and showed in Fig. \ref{fig:2} the
exemplified plots of the $(p,\mu)$ dependence of $F_\theta$ and
$F_\varphi$. From these plots one can see that they are symmetric
with respect to $p = 0.5$. In the region of $p\leqslant 0.5$, they
decrease monotonically with the increase of $p$. This phenomenon is
similar to the QFI for the depolarizing channel, and is also
understandable as the parameters in Eq. \eqref{eq3b-1} can be
obtained by replacing $\eta= 2p/3$ in Eq. \eqref{eq3a-2} with
$\eta=p$. However, due to the two extra nonzero elements
$\rho_{23,32}$, the resulting $\mu$ dependence of $F_\theta$ and
$F_\varphi$ are very different from those for the depolarizing
channel. As was shown in Fig. \ref{fig:2}, both $F_\theta$ and
$F_\varphi$ were slightly decreased by the classical correlations
between consecutive actions of the bit flip channel. But as their
dependence on $\mu$ is weak, such a detrimental effect is also very
weak.

For the same two qubits traversing the bit-phase flip channel, the
density operator of the output state has a similar form to that for
the bit flip channel. The only difference is that the matrix elements
$\rho_{23,32}$ are multiplied by a minus. One can then show that the
resulting $F_\theta$ and $F_\varphi$ have completely the same form to
those for the bit flip channel.

\subsection{The correlated phase flip channel} \label{sec:3c}
Now, we consider the case of two qubits traversing the correlated
phase flip channel, for which the nonzero elements of the density
operator for the output state are given by
\begin{equation}\label{eq3c-1}
 \begin{aligned}
  & \rho_{11}= \cos^2 \theta,~
    \rho_{44}=\sin^2 \theta, \\
  & \rho_{14}=\rho_{41}^*=w e^{-i\varphi} \sin\theta \cos\theta,
 \end{aligned}
\end{equation}
where $w=1-4p(1-p)(1-\mu)$.

For this case, one can also derive analytically its eigenvalues
$\{\lambda_i\}$, its eigenvectors $\{|\psi_i\rangle\}$, as well as
their partial derivatives and the resulting QFI. In fact, by combing
Eqs. \eqref{eq3a-1} and \eqref{eq3c-1}, one can see that the
corresponding QFI can be obtained directly by making the
substitutions $A\rightarrow 1$, $(B,C,E)\rightarrow 0$, and
$D\rightarrow w$ to those results showed in Appendix \ref{sec:A}.

\begin{figure}[h]
\centering
\resizebox{0.45 \textwidth}{!}{%
\includegraphics{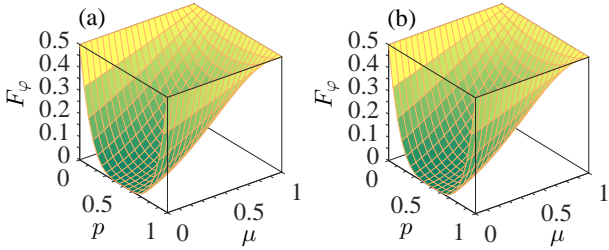}}
\caption{The $(p,\mu)$ dependence of $F_\varphi$ for the input state
$|\Phi^+\rangle$ and the correlated phase flip channel. The
parameters are $(\theta,\varphi)=(\pi/8,\pi/6)$ (a) and
$(\pi/8,\pi/3)$ (b).} \label{fig:3}
\end{figure}

Our calculation shows that no matter what the strengths of $p$ and
$\mu$ are, we always have $F_\theta=4$, i.e., it is immune of the
phase flip channel. So we showed in Fig. \ref{fig:3} only the
$(p,\mu)$ dependence of $F_{\varphi}$, which is symmetric with
respect to $p=0.5$. This is in fact an immediate consequence of the
form of $w$ below Eq. \eqref{eq3c-1}. In the region of $p\leqslant
0.5$, $F_{\varphi}$ decreases monotonically with the increasing
strength of $p$, while for any fixed $p\neq \{0,1\}$, $F_\varphi$
behaves as a monotonic increasing function of $\mu$ and reaches its
maximum value $F_\varphi^{\max}= \sin^2 {2\theta}$ when $\mu=1$.
This implies that the correlations between consecutive actions of
the phase flip channel are always beneficial for improving QFI of
the two-qubit state.

\begin{figure}[h]
\centering
\resizebox{0.46 \textwidth}{!}{%
\includegraphics{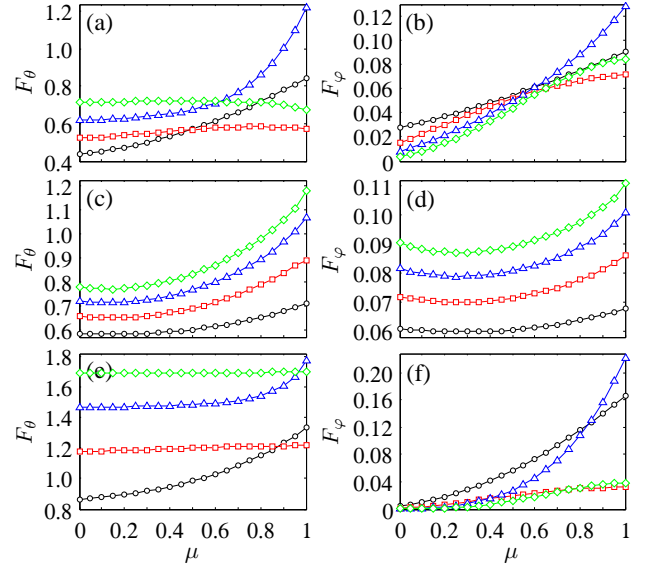}}
\caption{The $\mu$ dependence of $F_\theta$ and $F_\varphi$ for the
input state $\rho_0^{EWL}$ with $p=0.3$, $(\theta,\varphi)=
(\pi/8,\pi/6)$, $N=2$ (black circles), 3 (red squares), 4 (blue
triangles), and 5 (green diamonds). The top two, middle two, and
bottom two panels are plotted for the correlated depolarizing, bit
flip, and phase flip channels, respectively.}\label{fig:4}
\end{figure}

We have also performed calculations for other Bell-type probe states
including
\begin{equation}\label{eq3-2}
 \begin{aligned}
  & |\Phi^-\rangle= \cos\theta|00\rangle- e^{i\varphi}\sin\theta |11\rangle, \\
  & |\Psi^{\pm}\rangle= \cos{\theta}|01\rangle\pm e^{i\varphi}\sin{\theta}|10\rangle,
 \end{aligned}
\end{equation}
and found that $F_\theta$ and $F_\varphi$ show qualitatively the
same parameter dependence to those for the input state
$|\Phi^+\rangle$. This confirms again our observation that the
existence of classical correlations between consecutive actions of
the quantum channel on a sequence of qubits are beneficial for
improving the QFI. One may wonder whether the advantage of the
correlated quantum channels on improving QFI can be remained for
more general parameterization form of the probe states and for the
probe states with different number of qubit. To give an answer to
this concern, we further consider the following extended Werner-like
probe state
\begin{equation}\label{eq3-3}
 \rho_0^{EWL}= r|\Xi\rangle\langle\Xi|+\frac{1-r}{2^N}\openone_{2^N},
\end{equation}
where $|\Xi\rangle=\cos\theta|0\rangle^{\otimes N}+ e^{i\varphi}
\sin\theta |1\rangle^{\otimes N}$, and $r$ represents the ratio of
the component $|\Xi\rangle$ in $\rho_0^{EWL}$. In Fig. \ref{fig:4},
we presented an exemplified plot of the $\mu$ dependence of
$F_\theta$ and $F_\varphi$ for $\rho_0^{EWL}$ with different qubit
number $N$. It is clearly seen that the classical correlations
between consecutive actions of the quantum channel give an improved
QFI in a wide regime, hence the estimation precision of $\theta$ and
$\varphi$ can also be improved evidently. For certain cases, the
probe states with large number of qubit leads to a further
improvement of the QFI compared with that with small number of
qubit. This shows that the improved QFI due to correlated actions of
a quantum channel may applies to very general input probe states.

Finally, we present an interpretation on the different behaviors of
QFI for different quantum channels. Physically, such a difference is
due to the interplay between correlations arising from the
consecutive actions of the quantum channels on a sequence of qubits
and the temporal correlations occurring in the evolution of each
qubit. In particular, the temporal correlations are different for
different quantum channels. Moreover, the correlated actions of a
quantum channel leads to an inflow of information to the system
\cite{cc1,cc2,cc3}, and this explains the improved QFI. As QFI is
related to the variance of the quantum metrology, this further gives
an improved estimation precision of the physical parameters.

\section{Summary} \label{sec:4}
In summary, we have explored effects of the classical correlations
between consecutive actions of the quantum channel on QFI. Here, the
classical correlations were characterized by the probability that
the same Pauli transformation is applied to the sequence of qubits.
We have considered the depolarizing, bit flip, bit-phase flip, and
phase flip channels. For Bell-type probe states, it was found that
the decoherence effect is always detrimental to QFI. On the other
hand, effects of the classical correlations on QFI are different for
different quantum channels: (\romannumeral+1) For the correlated
depolarizing channel, by increasing the correlation strength $\mu$
from 0 to 1, $F_\theta$ first decreases to a certain minimum and
then turns to be increased to a finite value larger than that at
$\mu=0$, while $F_\varphi$ always behaves as an increasing function
of $\mu$. (\romannumeral+2) For the correlated bit flip and
bit-phase flip channels, both $F_\theta$ and $F_\varphi$ are
decreased slightly by increasing strength $\mu$ of the classical
correlations. (\romannumeral+3) For the correlated dephasing
channel, while $F_\theta$ keeps the constant 4, $F_\varphi$ can
always be increased by increasing $\mu$. In particular, it reaches
the maximum $\sin^2 {2\theta}$ for the limiting case of $\mu=1$,
irrespective of the strength of the decoherence parameter $p$.
For the more general parameterization form of the probe states, it
was found that the advantage of the correlated quantum channels on
improving QFI can also be remained. We hope these results may
deepen our understanding about role of the quantum channels on QFI
when they act on the sequence of qubits in different manners. They
are also expected to be helpful for constructing efficient schemes
to delay the rapid decay of QFI for open quantum systems.

\section*{ACKNOWLEDGMENTS}
This work was supported by NSFC (Grant No. 11675129), the New Star
Team of XUPT, and the Innovation Fund for Graduates (Grant No.
CXJJLY2018055).

\begin{appendix}
\begin{widetext}

\section{Derivation of QFI for the depolarizing channel} \label{sec:A}
\setcounter{equation}{0}
\renewcommand{\theequation}{A\arabic{equation}}

For the density operator $\rho$ of Eq. \eqref{eq3a-1}, its
eigenvalues are given by
\begin{equation}\label{eqa-1}
 \lambda_{1,2}= \frac{A+C\mp \alpha_1}{2},~
 \lambda_{3,4}=B,
\end{equation}
while the corresponding eigenvectors are given by
\begin{equation}\label{eqa-2}
 |\psi_{1,2}\rangle= \frac{(D+E)[(A-C)\cos 2\theta \mp \alpha_1] }
                     {(D e^{i\varphi}+Ee^{-i\varphi})\beta_{1,2}}|00\rangle
                     +\frac{(D+E)\sin 2\theta}{\beta_{1,2}} |11\rangle,~
 |\psi_3\rangle= |01\rangle,~
 |\psi_4\rangle= |10\rangle,
\end{equation}
where
\begin{equation}\label{eqa-3}
 \alpha_1=\sqrt{[(A-C)\cos 2\theta]^{2}+ [(D+E) \sin 2\theta]^{2}},~
 \beta_{1,2}=\sqrt{{2\alpha_1^{2}\mp 2(A-C)\alpha_1 \cos 2\theta}}.
\end{equation}

Then the partial derivatives of $\lambda_i$ and
$|\psi_i\rangle$ with respect to the parameter $\theta$ can be
obtained respectively as
\begin{equation}\label{eqa-4}
  \frac{\partial\lambda_{1,2}}{\partial\theta}=\pm\frac{[(A-C)^{2}-(D+E)^{2}]\sin4\theta}{2\alpha_1},~
  \frac{\partial\lambda_{3,4}}{\partial\theta}=0,
\end{equation}
and
\begin{equation}\label{eqa-5}
 \frac{\partial|\psi_{1,2}\rangle}{\partial\theta}= \frac{(D+E)\{h_{1,2}\beta_{1,2}
       \pm\delta_{1,2}[\alpha_1\pm(C-A)\cos2\theta]\}}{(D e^{i\varphi}+E e^{-i\varphi})\beta_{1,2}^{2}}|00\rangle
       +\frac{(D+E)(2\beta_{1,2}\cos2\theta-\delta_{1,2}\sin2\theta)}{\beta_{1,2}^{2}}|11\rangle,~
 \frac{\partial|\psi_{3,4}\rangle}{\partial\theta}=0,~
\end{equation}
where
\begin{equation}\label{eqa-6}
\begin{aligned}
 & h_{1,2}=\frac{2(C-A)\alpha_1 \sin2\theta\pm [(A-C)^{2}-(D+E)^{2}]\sin4\theta}{\alpha_1}, \\
 & \delta_{1,2}=\frac{[2\alpha_1\pm (C-A)\cos2\theta][(D+E)^{2}-(A-C)^{2}]\sin4\theta \pm 2(A-C)\alpha_1^{2}\sin2\theta}{\alpha_1\beta_{1,2}}.
\end{aligned}
\end{equation}

Similarly, the partial derivatives of $\lambda_i$ and $|\psi_i\rangle$
with respect to the parameter $\varphi$ can be obtained respectively
as $\partial \lambda_i/\partial \varphi =0$ ($\forall i$) and
\begin{equation}\label{eqa-7}
 \frac{\partial|\psi_{1,2}\rangle}{\partial\varphi}=\frac{-i(D+E)[(A-C)\cos2\theta
                \mp \alpha_1](D e^{i\varphi}-E e^{-i\varphi})}
                {(D e^{i\varphi}+E e^{-i\varphi})^{2}\beta_{1,2}}|00\rangle,~
 \frac{\partial|\psi_{3,4}\rangle}{\partial\varphi}=0.
\end{equation}

By substituting the above formulae into Eq. \eqref{eq2a-2}, one can obtain
$F_{\theta}$ and $F_{\varphi}$. As such substitutions are
straightforward and the resulting expressions are so complicated, we
do not list them here for concise of the presentation.

\section{Derivation of QFI for the bit flip channel} \label{sec:B}
\setcounter{equation}{0}
\renewcommand{\theequation}{B\arabic{equation}}

For the density operator $\rho$ of Eq. \eqref{eq3b-1}, its eigenvalues can
be obtained analytically as
\begin{equation}\label{eqb-1}
 \lambda_{1,2}=y(1\pm\sin2\theta \cos\varphi),~
 \lambda_{3,4}=\frac{x+z\mp \alpha_2}{2},
\end{equation}
while the corresponding eigenvectors are given by
\begin{equation}\label{eqb-2}
 |\psi_{1,2}\rangle=\frac{1}{\sqrt{2}}(|10\rangle\pm |01\rangle),~
 |\psi_{3,4}\rangle= {\frac{(x+z)[(x-z)\cos2\theta\mp \alpha_2]}{(xe^{i\varphi}
                     +ze^{-i\varphi})\beta_{3,4}}|00\rangle
                     +\frac{(x+z)\sin2\theta}{\beta_{3,4}}|11\rangle},
\end{equation}
where
\begin{equation}\label{eqb-3}
 \alpha_2=\sqrt{[(x-z)\cos2\theta]^{2}+[(x+z)\sin2\theta]^{2}}, ~
 \beta_{3,4}=\sqrt{2\alpha_2^{2}\pm 2(z-x)\alpha_2\cos2\theta}.
\end{equation}

Then the partial derivatives of $\lambda_i$ with respect to
$\theta$ can be obtained as
\begin{equation}\label{eqb-4}
 \frac{\partial\lambda_{1,2}}{\partial\theta}=\pm 2y\cos2\theta\cos\varphi,~
 \frac{\partial\lambda_{3,4}}{\partial\theta}=\mp\frac{2xz \sin4\theta}{\alpha_2},
\end{equation}
while the partial derivatives of $|\psi_i\rangle$ with respect to
$\theta$ can be obtained as
\begin{equation}\label{eqb-5}
 \frac{\partial|\psi_{1,2}\rangle}{\partial\theta} =0,~
 \frac{\partial|\psi_{3,4}\rangle}{\partial\theta}= \frac{(x+z)\{h_{3,4}\beta_{3,4}
       +\delta_{3,4}[(z-x)\cos2\theta\pm \alpha_2]\}}{(xe^{i\varphi}+ze^{-i\varphi})\beta_{3,4}^{2}}|00\rangle
       +\frac{(x+z)(2\beta_{3,4}\cos2\theta-\delta_{3,4}\sin2\theta)}{\beta_{3,4}^{2}}|11\rangle,
\end{equation}
where
\begin{equation}\label{eqb-6}
  h_{3,4}= \frac{2(z-x)\alpha_2\sin2\theta+ 4xz \sin4\theta}{\alpha_2},~
  \delta_{3,4}= \frac{4xz[2\alpha_2\pm (z-x)\cos2\theta]\sin 4\theta
                \pm 2(x-z)\alpha_2^{2}\sin2\theta}{\alpha_2\beta_{3,4}}.
\end{equation}

Next, the partial derivatives of $\lambda_i$ with respect to
$\varphi$ can be obtained as
\begin{equation}\label{eqb-7}
 \frac{\partial\lambda_{1,2}}{\partial\varphi}= \mp y\sin2\theta\sin\varphi,~
 \frac{\partial\lambda_{3,4}}{\partial\varphi}=0,
\end{equation}
and the partial derivatives of $|\psi_i\rangle$ with respect to
$\varphi$ can be obtained as
\begin{equation}\label{eqb-8}
\frac{\partial|\psi_{1,2}\rangle}{\partial\varphi} =0,~
\frac{\partial|\psi_{3,4}\rangle}{\partial\varphi}= \frac{i(x+z)[(z-x)\cos2\theta
                      \pm \alpha_2](xe^{i\varphi}-ze^{-i\varphi})}{(xe^{i\varphi}
                      + ze^{-i\varphi})^{2}\beta_{3,4}}|00\rangle.
\end{equation}

With these expressions on hand, one can obtain the corresponding
$F_\theta$ and $F_\varphi$ directly.

\end{widetext}
\end{appendix}

\newcommand{\PRL}{\emph{Phys. Rev. Lett.} }
\newcommand{\RMP}{\emph{Rev. Mod. Phys.} }
\newcommand{\PRA}{\emph{Phys. Rev. A} }
\newcommand{\PRB}{\emph{Phys. Rev. B} }
\newcommand{\PRE}{\emph{Phys. Rev. E} }
\newcommand{\NJP}{\emph{New J. Phys.} }
\newcommand{\JPA}{\emph{J. Phys. A} }
\newcommand{\JPB}{\emph{J. Phys. B} }
\newcommand{\OC}{\emph{Opt. Commun.} }
\newcommand{\PLA}{\emph{Phys. Lett. A} }
\newcommand{\EPJD}{\emph{Eur. Phys. J. D} }
\newcommand{\NP}{\emph{Nat. Phys.} }
\newcommand{\NPo}{\emph{Nat. Photonics} }
\newcommand{\NC}{\emph{Nat. Commun.} }
\newcommand{\EPL}{\emph{Europhys. Lett.} }
\newcommand{\AoP}{\emph{Ann. Phys.} }
\newcommand{\QIC}{\emph{Quantum Inf. Comput.} }
\newcommand{\QIP}{\emph{Quantum Inf. Process.} }
\newcommand{\CPB}{\emph{Chin. Phys.} B }
\newcommand{\IJTP}{\emph{Int. J. Theor. Phys.} }
\newcommand{\IJQI}{\emph{Int. J. Quantum Inf.} }
\newcommand{\IJMPB}{\emph{Int. J. Mod. Phys. B} }
\newcommand{\PR}{\emph{Phys. Rep.} }
\newcommand{\SR}{\emph{Sci. Rep.} }
\newcommand{\LPL}{\emph{Laser Phys. Lett.} }
\newcommand{\SCG}{\emph{Sci. China Ser. G} }
\newcommand{\JSP}{\emph{J. Statis. Phys.} }


\end{document}